\documentclass[aps,prb,twocolumn,floatfix]{revtex4}

\usepackage{graphicx}

\newcommand{\ddt}[1]{{{\partial {#1}\over \partial t}}}
\newcommand{\ddx}[1]{{{\partial {#1}\over \partial x}}}

\newcommand{\ddxx}[1]{{{\partial^2 {#1}\over \partial x^2}}}

\newcommand{\bb}{{{\bf b}}}

\newcommand{\kk}{{{\bf k}}}

\newcommand{\ttt}{{{\bf t}}}

\newcommand{\vv}{{{\bf v}}}

\newcommand{\rot}{{{\nabla \times}}}
\newcommand{\divr}{{{\nabla \cdot}}}
\newcommand{\inv}[1]{{{1\over #1}}}

\newcommand{\ie}{{{\it i.e.}}}
\newcommand{\etal}{{{\it et. al.}}}

\begin{document}

\preprint{npg-2006-0042}

\title{Remarks on nonlinear relation among phases and frequencies
in modulational instabilities of parallel propagating Alfv\'en waves}

\author{Y. Nariyuki}
 \altaffiliation[Also at ]{Department of Earth System Science 
 and Technology, Kyushu University}
\author{T. Hada}%


\date{\today}

\begin{abstract}
Nonlinear relations among frequencies and phases in modulational instability of circularly polarized Alfv\'en waves are discussed, within the context of one dimensional, dissipation-less, unforced fluid system. We show that generation of phase coherence is a natural consequence of the modulational instability of Alfv\'en waves. Furthermore, we quantitatively evaluate intensity of wave-wave interaction by using bi-coherence, and also by computing energy flow among wave modes, and demonstrate that the energy flow is directly related to the phase coherence generation. We first discuss the modulational instability within the derivative  nonlinear Schr\"odinger (DNLS) equation, which is a subset of the Hall-MHD system including the right- and left-hand  polarized, nearly degenerate quasi-parallel Alfv\'en waves. The dominant nonlinear process within this model is the four wave interaction, in which a quartet of waves in resonance can exchange energy. By numerically time integrating the DNLS equation with periodic boundary conditions, and by evaluating relative phase among the quartet of waves, we show that the phase coherence is generated when the waves exchange energy among the quartet of waves. As a result, coherent structures (solitons) appear in the real space, while in the phase space of the wave frequency and the wave number, the wave power is seen to be distributed around a straight line. The slope of the line corresponds to the propagation speed of the coherent structures. Numerical time integration of the Hall-MHD system with periodic boundary conditions reveals that, wave power of transverse modes and that of longitudinal modes are aligned with a single straight line in the dispersion relation phase space, suggesting that efficient exchange of energy among transverse and longitudinal wave modes is realized in the Hall-MHD. Generation of the longitudinal wave modes violates the assumptions employed in deriving the DNLS such as the quasi-static approximation, and thus long time evolution of the Alfv\'en modulational instability in the DNLS and in the Hall-MHD models differs significantly, even though the initial plasma and parent wave parameters are chosen in such a way that the modulational instability is the most dominant instability among various parametric instabilities. One of the most important features which only appears in the Hall-MHD model is the generation of sound waves driven by ponderomotive density fluctuations. We discuss relationship between the dispersion relation, energy exchange among wave modes, and coherence of phases in the waveforms in the real space. Some relevant future issues are discussed as well. 

(accepted to Nonlinear Processes in Geophysics)
\end{abstract}

\pacs{Valid PACS appear here}
\maketitle

%
\section{Introduction}

In various areas in space and astrophysical environment, for instance in
the solar wind and in foreshock region of planetary bowshocks,
parametric instabilities of magnetohydrodynamic (MHD) waves are
thought to play essential roles in generating MHD turbulence.
Spacecraft observations suggest that, within the MHD turbulence
in the solar wind and in the earth's foreshock,
a large number of localized structures are often embedded
(Mann \etal, 1994; Dudok de Wit \etal, 1999; Lucek \etal, 2004; 
Tsurutani \etal, 2005).
Also, recent surrogate data analysis using magnetic field data obtained by
Geotail spacecraft (Hada \etal, 2003; Koga and Hada, 2003) has 
revealed that
large amplitude MHD waves observed in the earth's foreshock are not
completely phase random, but are almost always phase correlated to a 
certain degree.
Furthermore, the larger the MHD wave amplitude, the stronger the wave phase
correlation, implying that the detected phase coherence is a 
consequence of nonlinear
interaction among the MHD waves.
Since the finite phase coherence in the Fourier space corresponds 
to the presence
of localized structures in the real space, the large amplitude MHD 
turbulence in the
foreshock should be regarded as a superposition of random phase
MHD turbulence plus a large number of localized structures.

Since the MHD waves are dispersive in general, it is natural to infer 
that the localized
structures once produced would disperse away as time elapses.
On the other hand, the nonlinearity of the MHD set of equations acts 
as a source for production of the localized structures.
In fact, inspection of a simple model representing interaction of 
parallel Alfv\'en wave modes suggests that, the phase coherence is
generated whenever there is an exchange of wave energy among wave modes
in resonance (Nariyuki and Hada, 2005).
Namely, the localized structures are unstable and fade away, but are also
continuously born due to intrinsic nonlinearity of the MHD system.
Such behavior of the localized structures is typically seen in
nonlinear evolution of Alfv\'en wave parametric instabilities,
in particular, in nonlinear evolution of the modulational instability.
In a similar context, Nocera and Buti (1996) discussed formation of 
localized pulses in a non-dissipative DNLS system, and Hasegawa \etal (1981) 
observed emergence of organized structures in the Korteweg - de Vries equation.
The four-wave interaction scheme was used to explain the emergence of
organized structures in the unforced, dissipative DNLS (Krishan and Nocera, 2003).
In this paper, we discuss implications of the presence of finite phase
correlation, which corresponds to the presence of localized structures,
generated by the wave-wave interactions in modulational instability of
circularly polarized Alfv\'en waves, within the context of one dimensional, 
dissipation-less, unforced fluid system.
We will emphasize the importance of nonlinear relations among
frequencies and phases in identifying relevant physical process
involved in the modulational instability. 

The plan of the paper is as follows:
In Section 2, we review and compare the parametric instabilities
in the Hall-MHD system and in the DNLS equation.
In Section 3, we discuss the relationship between generation of solitary waves
and nonlinear wave-wave interaction in detail,
by examining numerically produced Alfv\'enic turbulence using the 
DNLS equation.
Our main discussions on the phase coherence is presented in this section.
In Section 4, we discuss phase and frequency features in the Hall-MHD system.
From the relation between the plasma density and the envelope of the 
magnetic field,
we examine validity of the so-called quasi-static approximation 
used in the DNLS.
When the longitudinal and transverse wave modes have similar phase velocities,
they can couple strongly, and as a result the phase coherence can 
be generated.
We summarize the results and discuss some of the fundamental issues presented
in the paper in Section 5. 

%
%
\section{Basic equations and analytical study of parametric instabilities}

It has long been known that circularly polarized ('parent') Alfv\'en waves are
parametrically unstable to generation of plasma density fluctuations
and 'daughter' Alfv\'en waves with the same polarization as the 'parent'.
As early as in the 1960's, it was shown by Galeev and Oraevskii (1963), 
and Sagdeev and Galeev (1969) that the circularly polarized Alfv\'en waves 
are subject to parametric decay instability (see Appendix).
Later, Goldstein (1978) and Derby (1978) derived the dispersion relation of
the decay instability of the circularly polarized Alfv\'en waves using 
ideal MHD equation set.
Roles of the dispersion effect was investigated using the Hall-MHD 
(two-fluid) set of
equations by Wong and Goldstein (1986), Longtin and Sonnerup (1986), and
Terasawa \etal (1986). 
The Hall-MHD equations are
\begin{eqnarray}
\ddt{\rho} &=& -\divr (\rho \vv)\\
\ddt{\vv} &=& - \vv\cdot\nabla \vv - \inv{\rho}\nabla
\left( p + {|\bb|^2 \over 2} \right) + \inv{\rho} (\bb\cdot\nabla)\bb\\
\ddt{\bb} &=& \rot (\vv\times\bb) - \rot
\left( \inv{\rho} (\rot\bb)\times\bb \right) \\
\divr\bb &=& 0
\end{eqnarray}
where
the density $\rho$ is normalized to the initial uniform density $\rho_0$,
the magnetic field $\bb$ to the background constant field magnitude $b_{0x}$,
the velocity $\vv$ to the Alfv\'en velocity defined
by $\rho_0$ and $b_{0x}$,
and the pressure $p$ to the ambient magnetic pressure.
Time and space are respectively normalized to the reciprocal of the 
ion cyclotron frequency and the ion inertial length defined 
using the background quantities.

In this paper, all the physical quantities are assumed to be dependent only on
one spatial coordinate ($x$).
Then the governing equations become
\begin{eqnarray}
\ddt{\rho} &=& - \ddx{}(\rho u)\\
\ddt{u} &=& - u \ddx{u} - \inv{\rho}\ddx{}\left( p + {|b|^2\over 2} \right)\\
\ddt{v} &=& -u \ddx{v} + \inv{\rho} \ddx{b}\\
\ddt{b} &=& - \ddx{} \left( u b - v + {i \over \rho} \ddx{b} \right)
\end{eqnarray}
where $b=b_y + ib_z$ and $v=v_y + iv_z$ are the complex transverse
magnetic field and velocity, respectively,
and $u=v_x$ is the longitudinal velocity.
For simplicity, we assume the equation of state to be isothermal, 
$ p = \beta \rho$,
where a constant $\beta$ is the squared normalized sound wave speed.

In the following, linear perturbation analysis is used to discuss 
parametric instabilities
of parallel propagating Alfv\'en waves, using the DNLS equation 
(Section 2.1) and
the Hall-MHD equations (Section 2.2).

\subsection{Linear perturbation analysis of the DNLS system}

By applying a quasi-static approximation in which hydrodynamic
nonlinearities and steeping are weak
(\ie, variation of the plasma density is caused only by magnetic ponderomotive
fluctuations), Rogister (1971) derived a kinetic equation describing 
the long time
evolution of the Alfv\'en waves.
Essentially the same equation was subsequently obtained starting from 
the two-fluid set
of equations (Mj{\o}lhus, 1974; Mio \etal, 1976; Spangler and Sheerin, 1982;
Sakai and Sonnerup, 1983; Mj{\o}lhus and Hada, 1997).
The fluid version, now known as the DNLS, reads
\begin{equation}
\ddt{b} + \alpha \ddx{}(|b|^2 b) + i \mu \ddxx{b} = 0 ,
\end{equation}
where normalizations are the same as in (1)-(4),
$\alpha / C_A =C_i^2 / 4(C_i^2 - C_s^2)$,
$2\mu/C_A =c/\omega_{pi}$ is the ion inertia dispersion length,
and
$C_A$, $C_i$, $C_s$ are the Alfv\'en, intermediate, and sound
speeds, respectively.
In the DNLS equation, $b=-v$, and
\begin{equation}
u=\rho=\frac{|b|^{2}}{2(1-\beta)} ,
\end{equation}
where $\beta = C_s^2/C_A^2$ (the quasi-static approximation).

The DNLS describes evolution of weakly nonlinear, quasi-parallel
propagating, both right- and left-hand polarized Alfv\'en waves
(or "magnetosonic" and "shear Alfv\'en" waves),
which are nearly degenerate.
Modulational instability is driven unstable for the left-hand
polarized waves when $\beta < 1$,
and for the right-hand polarized waves when $\beta>1$
(Mj{\o}lhus, 1976; Spangler and Sheerin, 1982; Sakai and Sonnerup, 1983),
although presence of resonant ions significantly alter the
above conditions, especially for $\beta>1$ and moderate to large
ion to electron temperature ratio
(Rogister, 1971; Mj{\o}lhus and Wyller, 1988; Spangler, 1990;
Medvedev \etal, 1997).
The DNLS is known to be integrable under various boundary
conditions (Kaup and Newell, 1978; Kawata and Inoue, 1978;
Kawata \etal, 1980; Chen and Lam, 2004).
In this paper, we restrict our discussion to the case
$\beta \ll 1$, since only within this regime
the use of the fluid version of the DNLS is justified, unless the ion 
to electron temperature ratio is extremely low.

By re-scaling of the variables,
$t\rightarrow \mu t/\alpha^2$ and $x \rightarrow \mu x/\alpha$,
(9) is reduced as
\begin{equation}
\ddt{b} + \ddx{}(|b|^2 b) + i \ddxx{b} = 0 .
\end{equation}
A parallel, circularly polarized, finite amplitude Alfv\'en wave,
$b_p = b_0 \exp(i(\omega_0 t -k_0 x))$,
where $b_0$, $\omega_0$, and $k_0$ are real, 
satisfies (11) with the dispersion relation for the
parent wave, $\omega_0 = b_0^{2} k_0 + k_0^{2}$. 
Now we superpose small fluctuations to $b_p$ and write
\begin{equation}
b =b_p+\sum_{n=-\infty}^{\infty}
\epsilon^{|n|}b_n\exp(i( \omega_{n} t - k_{n}x )) ,
\end{equation}
where $k_{n}=k_0 + nK$,
$\omega_{n}=\omega_0 + n\Omega$, and $n$ is an integer (except for 0).
Here, $\Omega$ and $K$ are the frequency and the wave number of
longitudinal perturbations (quasi-modes), $u=\rho$.
At order of $\epsilon^{|\pm 1|}$, we obtain linearized equations
\begin{equation}
\Omega = 2K(b_0^2+k_0) \pm K \sqrt{K^2+2b_0^2k_0+b_0^4} .
\end{equation}
If $0<|K|<b_0\sqrt{-2k_0-b_0^2}$, the system exhibits the 
modulational instability.
The wave number corresponding to the maximum growth rate is
$K_{max} = b_0 (-k_0-b_0^2/2)^{1/2}$.
When the system is modulationally unstable, 
the frequency of sideband modes is obtained from the real part of (13) as
$\omega_{\pm 1} = 2 v_{\phi 0} k_{\pm 1} - \omega_0$,
where $v_{\phi 0} = \omega_{0}/k_{0}$.
Therefore, the dispersion relation of the sideband modes appears as
a straight line, since the dispersion term is cancelled by the nonlinear term.
In other words, the nonlinear effect and the dispersion effect are balanced.
We will encounter again such a dispersion relation in numerical analysis in 
Section 3 and 4,
where more detailed discussions will be given.

\subsection{Linear perturbation analysis of the Hall-MHD equation}

Finite amplitude, circularly polarized Alfv\'en wave in the form below
is an exact solution to (5)-(8),
\begin{eqnarray}
b_p &=& b_0 \exp(i(\omega_0 t -k_0 x)) ,\\
v_p &=& v_0 \exp(i(\omega_0 t -k_0 x)) ,
\end{eqnarray}
where $v_0 = -b_{0}/v_{\phi 0}$, $v_{\phi 0}=\omega_{0}/k_{0}$,
$\rho=1$, and $u=0$, together with the dispersion relation
\begin{equation}
\omega_0^2 = k_0^2 (1+\omega_0) .
\end{equation}
Now we add small fluctuations and write
\begin{eqnarray}
\rho &=&1+{1\over 2}\sum_{n=1}^{\infty}
\epsilon^{|n|}\rho_n\exp(i( n \Omega t - n K x )) + c.c.,\\
u &=&0+{1\over 2}\sum_{n=1}^{\infty}
\epsilon^{|n|}u_n\exp(i( n \Omega t - n K x )) + c.c.,\\
v &=&v_p+\sum_{n=-\infty}^{\infty}
\epsilon^{|n|}v_n\exp(i( \omega_{n} t - k_{n}x )) ,\\
b &=&b_p+\sum_{n=-\infty}^{\infty}
\epsilon^{|n|}b_n\exp(i( \omega_{n} t - k_{n}x )) ,
\end{eqnarray}
where $k_{n}=k_0 + n K$, $\omega_{n}=\omega_0 + n\Omega$, $n$ is an integer, 
and $c.c.$ represents the complex conjugate. 
At order of $\epsilon^{|\pm 1|}$, we obtain the following linearized equations
\begin{eqnarray}
\Omega \rho_1 &=& K u_1 , \\
\Omega u_1    &=& K b_0 (b_{+} + b_{-}^*) + \beta K \rho_{1} ,\\
\omega_{+} b_{+}   &=& -k_{+} v_{+} + k_{+} \frac{b_0 u_1}{2} + k_{+}^2 b_{+}
- \frac{b_0 k_0 k_{+}}{2} \rho_{1} ,\\
\omega_{-} b_{-}^* &=& -k_{-} v_{-}^* + k_{-} \frac{b_0 u_1}{2} + k_{-}^2 b_{-}^*
- \frac{b_0 k_0 k_{-}}{2} \rho_{1} ,\\
\omega_{+} v_{+}   &=&  \frac{k_0 v_0}{2} u_{1} - k_{+} b_{+}   +\frac{b_0 k_0}{2} \rho_{1} ,\\
\omega_{-} v_{-}^* &=&  \frac{k_0 v_0}{2} u_{1} - k_{-} b_{-}^* +\frac{b_0 k_0}{2} \rho_{1} ,
\end{eqnarray}
where we write the subscripts $(\pm 1)$ as $(\pm)$ for brevity.
Combining these equations above, we obtain,
\begin{equation}
(\Omega^2 - \beta K^{2}) L_{+} L_{-} = \frac{K^{2} b_0^2}{2} (L_{+} R_{-} 
+ L_{-} R_{+}) ,
\end{equation}
where
\begin{equation}
L_{\pm}=\omega_{\pm}^2 -k_{\pm}^2(1+\omega_{\pm}) ,
\end{equation}
\begin{equation}
R_{\pm}=k_0 k_{\pm} \left( 
\frac{k_{0}\Omega}{\omega_{0}K}+\frac{\Omega\omega_{\pm}}{ k_{0}K} 
-1-\omega_{\pm} \right) .
\end{equation}
%
%
\begin{figure}[bt]
\begin{center}
\includegraphics[width=8cm]{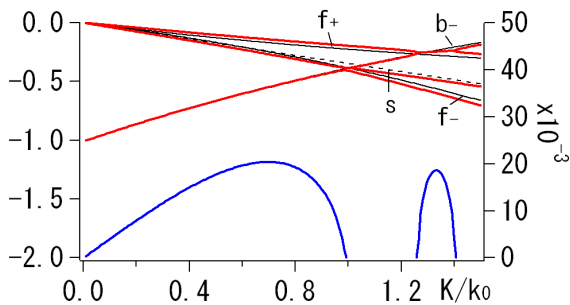}
\end{center}
\caption{
Dispersion relation of parametric instabilities in the 
Hall-MHD system. Shown are the real (red lines, left scale)
and imaginary (blue lines, right scale) normalized frequencies plotted
versus the normalized wave number, $K/k_0$.
Parameters used are, $k_0=-0.5$, $\beta = 0.5$, and $b_0 = 0.4$.
For comparison, the dispersion relation of sideband wave modes and 
the sound wave (zero parent wave amplitude) is superposed
as black solid and dashed lines, respectively. 
Forward and backward lower (upper) sideband waves are labeled
$f-(+)$ and $b-(+)$, and the sound wave is labeled $s$.}
\end{figure}

The dispersion relation (27) has been obtained and solved numerically
(Wong and Goldstein, 1986; Longtin and Sonnerup, 1986; Terasawa \etal, 1986;
Vinas and Goldstein, 1991; Hollweg, 1994; Champeaux \etal, 1999).
Fig. 1 shows the real and imaginary frequencies of the longitudinal waves
when $\beta=0.5$, $b_0=0.4$, and $k_0=0.5$.
Under this particular set of parameters, the modulational instability 
($0<K/k_0 < 1$) has a growth rate larger than (or almost comparable with)
the beat instability around $k=1.3$.

It is instructive to look at (23) and (24) in order to discuss how 
the side band mode frequencies are determined.
The r.h.s. of these equations represent three basic processes
which determine the transverse magnetic field;
the linear response ($ k_{\pm} v_{\pm}^{(*)}$), 
the ${\bf v} \times {\bf b}$ nonlinearity ($k_{\pm}b_{0}u_{1}/2$), 
and the Hall effect ($k_{\pm}^{2}b_{\pm}^{(*)}-b_{0}k_{0}k_{\pm}\rho /2$).
Hereafter we will call these terms as 
"linear", "nonlinear", and the "dispersive" terms, respectively.
For a given set of parent wave parameters ($b_0$ and $k_0$), $\beta$, 
and $K$, one can solve (27) to find $\Omega$ associated with the 
modulational instability (eigenvalue), together with ratios among 
variables, 
$\rho_1$, $u_1$, $b_{\pm}^{(*)}$, and $v_{\pm}^{(*)}$ (eigenvector).
By varying $K$ in both positive and negative regimes, the dispersion relation
can be visualized by plotting the real part of $\omega \equiv \omega_0 + \Omega$
as a function of $k \equiv k_0 + K$, as shown in Fig. 2 as a solid red line. 
Superposed in the figure are the plots of $ \omega_{lin} = (k_{\pm} v_{\pm}^{(*)})/b_{\pm}^{(*)}$, 
$\omega_{nl} = k_{\pm}b_{0}u_{1}/2 b_{\pm}^{(*)} $, and 
$\omega_{disp}= k_{\pm}^{2}-b_{0}k_{0}k_{\pm}\rho /2 b_{\pm}^{(*)}$ versus $k$, 
which are frequency contribution from the three effects introduced above. 
By definition, $\omega = \omega_{lin} + \omega_{nl} + \omega_{disp}$. 

From the figure we notice that 
both $\omega$ and $\omega_{lin}$ almost linearly depend on $k$, 
and thus sum of $\omega_{nl}$ and $\omega_{disp}$ should also depend
linearly on $k$, suggesting that the nonlinear effect and the 
dispersion effect are balanced, as they were for the DNLS equation
discussed previously. 
In other words, the mismatch of the resonance condition due to the 
dispersion effect
is reduced by nonlinear interaction among the Fourier modes,
just like in the case of the DNLS (Section 2.1).
%
%
\begin{figure}[bt]
\begin{center}
\includegraphics[width=8cm]{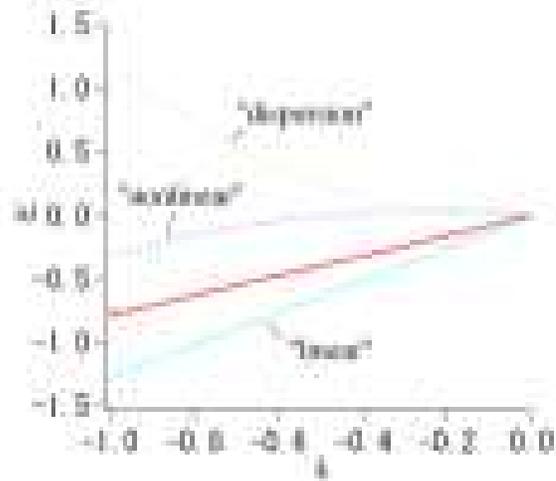}
\end{center}
\caption{
The three frequency contributions, 
$\omega_{lin}$, $\omega_{nl}$, and $\omega_{disp}$, 
plotted versus $k$, using the same set of parameters used for Fig. 1.
See text for explanation of how to make the plot. 
Both Re[$\omega$] (red solid line) and $\omega_{lin}$ are
linearly dependent on $k$, and so should be the sum of $\omega_{disp}$
and $\omega_{nl}$, suggesting that the dispersion and the 
nonlinear effects are cancelled each other. }
\end{figure}

%
%
\section{Phase coherence in the DNLS equation}

In this Section, we discuss the relationship between generation of 
solitary waves and
the wave-wave interaction.
In Fig. 3 we show numerical time integration of (11) under periodic
boundary conditions, where the envelope ($|b|$) is plotted in the
phase space of time ($t$) and space ($x$).
As initial conditions, finite amplitude, left-hand polarized,
monochromatic Alfv\'en waves with the wave amplitude $b_0=0.4$
and the wave mode number $m = m_0 = -11$ are given,
superposed with a very small amplitude white noise with
$<|b_{noise}|^2>^{1/2}=10^{-5}$ within the range of
$-256 < m < 256$, where the bracket denotes spatial average
over the simulation system.
In the above, the wave number $k$ is related to the mode number $m$
as $k = 2\pi m/L$, where the system size is $L=256$.
We have used the convention that $m$ and $k$ positive (negative)
represents the right- (left-) hand polarized waves.
For the numerical computation, we have employed 
the rationalized Runge-Kutta scheme for time integration and the 
spectral method for evaluating spatial derivatives.
Number of grid points used for this run is 2048.

Since in Fig. 3 we are plotting the magnetic field envelope,
which is constant for the monochromatic, circularly polarized waves,
at the beginning not much wave activities are apparent.
However, starting from $t \sim 200$, modulation of the envelope
becomes increasingly more evident, and around $t \sim 300$
a series of solitary waves is created.
The number of solitary waves is decided by the wave number that has the
largest growth rate in (13).
After $t \sim 300$ we observe a complex behavior of solitary waves:
they appear, propagate, disappear, and interact with each other.
%
%
\begin{figure}[bt]
\begin{center}
\includegraphics[width=8cm]{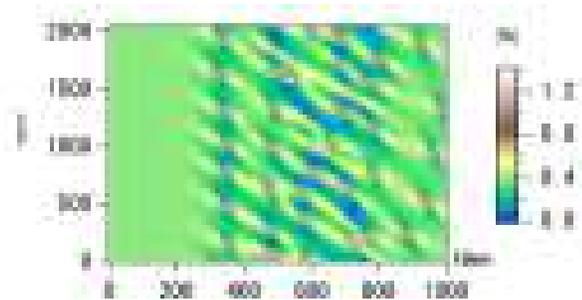}
\end{center}
\caption{Time evolution of envelope $|b|$ in the DNLS model (11)  
with periodic boundary conditions. 
Initial conditions are given as a superposition
of finite amplitude, left-hand polarized monochromatic waves,
and very small amplitude white noise.}
\end{figure}

Corresponding time evolution of the power spectrum is
shown in Fig. 4.
During $0 < t \lesssim 200$,
the parent wave energy ($m=-11$) is gradually transferred
to the side-band daughter waves (mainly, $m=-4$ and $m=-18$)
through the modulational instability.
Later on, increasingly more waves at different mode numbers
are generated due to coupling among finite amplitude waves,
and also due to the modulational instability of the daughter waves,
as can be seen as widening of the power spectrum in the $m$-space.
Around $t \sim 300$, the width of the spectrum is maximized,
corresponding to the appearance of the solitary waves.
When the solitary waves disappear, the power spectrum
becomes narrow again, representing the uncertainty principle,
\ie, the width of the wave packet in the real and Fourier space are
inversely proportional to each other.
We note that solitary waves disappear almost completely at
certain time intervals like $t \sim 900$.
This is reminiscent of the presence of (near) recursion of the DNLS equation
with periodic boundary conditions.
%
%
\begin{figure}[bt]
\begin{center}
\includegraphics[width=8cm]{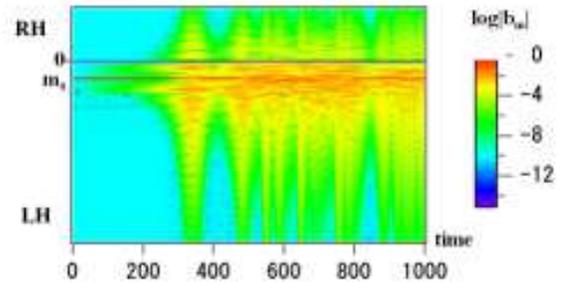}
\end{center}
\caption{Time evolution of the power spectrum (in logarithmic scale),
$\log|b_m|$, plotted in the phase space of the wave mode number ($m$)
and time.
The parent mode is given initially at $m_0=-11$.
Positive (negative) $m$ corresponds to
right- (left-) hand polarized waves.
The wave number is given by $k=2\pi m/L$, where $L=256$ is the system size.
}
\end{figure}

Next we show that distribution of wave phases and frequencies are 
closely related to
the generation and disappearance of the solitary waves.
We write the Fourier transformed magnetic field 
$b_k=|b_k|\exp{i\phi_k}$, and discuss
the relation between behavior of solitary waves and correlation among
wave phases, $\phi_k$.
A method to evaluate the wave phase coherence has been proposed
(Hada \etal, 2003; Koga and Hada, 2003), which we briefly explain below:
Suppose we have a data containing waves, $B_{ORG}(x)$, 
where 'ORG' stands for 'original data'.
In the present analysis, this is a snapshot of simulation data evaluated
at certain fixed time. 
From $B_{ORG}(x)$ we make a phase randomized surrogate (PRS), $B_{PRS}(x)$, 
and a phase correlated surrogate (PCS), $B_{PCS}(x)$, by
shuffling and making equal all the phases, respectively.
Then we compute the "phase coherence index",
\begin{equation}
C_{\phi}=(L_{PRS}-L_{ORG})/(L_{PRS}-L_{PCS}) ,
\end{equation}
where
\begin{equation}
L_{*}=\sum_{x}|B_{*}(x+\delta)-B_{*}(x)| 
\end{equation}
is the first order structure function evaluated for data (*), and
the asterisk stands for either ORG, PRS, or PCS.
In the above, $\delta$ is an external parameter representing the
coarsing scale, which we choose to be the grid size for the present analysis.
When $C_{\phi}$ is close to 0, wave phases are almost random,
while $C_{\phi}$ being close to unity suggests that phases are almost 
completely coherent.
Fig. 5 shows time evolution of $C_{\phi}$ for the run shown in Fig. 3.
Apparently, the appearance and disappearance of solitary wave trains
correspond to the increase and decrease of $C_{\phi}$.
This is easily understood since the solitary waveform may be
produced by equal-phase superposition of many waves with different wavelengths.

%
%
\begin{figure}[bt]
\begin{center}
\includegraphics[width=8cm]{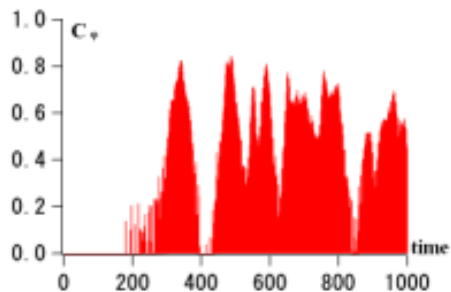}
\end{center}
\caption{Time evolution of the phase coherence index, $C_{\phi}$.
When $C_\phi \sim 0$, the wave phases are almost random,
while $C_\phi \sim 1$ represents that the wave phases are
almost completely correlated.
}
\end{figure}

Now we discuss how phase coherence is generated by the modulational 
instability in the DNLS.
Since the nonlinearity of the DNLS equation is cubic, it follows that,
the basic nonlinear interaction in this model is the four wave resonance.
(On the other hand, if we regard $|b|^2$ as a quasi-mode as represented 
in the static approximation (10), 
the interaction can be viewed as the three wave resonance
among two Alfv\'en waves and the quasi-mode.)

We have the resonance relation among the wave numbers,
\begin{equation}
k_1 + k_2 = k_3 + k_4 , 
\end{equation}
and a similar relation should hold for the wave frequencies as well.
While the resonance condition of the wave frequencies usually has the mismatch
due to the dispersion effect in the DNLS,
as we saw in Section 2, the dispersion effect and the nonlinear effect cancels
each other, and so there is no mismatch of frequencies when the wave modes
are coupled.

Let us define the relative phase among the four waves as
\begin{equation}
\theta (\kk) \equiv \phi_{k1}+\phi_{k2}-\phi_{k3}-\phi_{k4} ,
\end{equation}
where $\kk = \{ k_1, k_2, k_3, k_4 \}$ represents the quartets of wave modes
in resonance.
Small temporal change of $\theta(\kk)$ implies that the phase coherence
between the four waves is strong (locked), because this temporal 
change corresponds
to the mismatch of the resonance condition of frequencies.

From (11) and (33), we have
\begin{equation}
\frac{d}{dt}\left(\frac{|b_{k1}|^2}{2}\right) =
k_1\sum |b_{k1}||b_{k2}||b_{k3}||b_{k4}|\sin\theta (\kk) , 
\end{equation}
and
\begin{equation}
\frac{d}{dt}\left(\phi_{k1}\right) =
k_1\sum \frac{|b_{k2}||b_{k3}||b_{k4}|}{|b_{k1}|}\cos\theta (\kk)+k_{1}^{2} , 
\end{equation}
where the summation is to be taken over all the combinations of
$(k_2, k_3, k_4)$ which satisfy (32).
Equation (34) indicates that the evolution of wave energy of a
certain mode is determined by 'energy flow' exchanged between the quartets,
\begin{equation}
F (\kk) \equiv k_1|b_{k1}||b_{k2}||b_{k3}||b_{k4}|\sin\theta (\kk) .
\end{equation}
The direction of the energy flow is determined by the sign of
$\sin\theta(\kk)$ and $k_1$.

Fig. 6 is a scatter plot in the phase space of $F(\kk)$ 
and $|d\theta(\kk)/dt|$ ($=|\dot{\theta(\kk)}|$),
in which a single dot is plotted at every time step based on the 
simulation run shown in Fig. 3. 
The set of wavenumbers, $k_j = 2 \pi m_j / L$, with 
$m_3 = m_4 = m_0 = -11$, $m_1 = -15$, $m_2 = -7$ are chosen 
in such a way that the waves satisfy the resonance condition, 
and also that the modes $m=m_1$ and $m_2$ are the two daughter waves
driven unstable by the parent wave at $m=m_0$. 
It is seen that exchange of the wave energy among the four wave
modes is enhanced (reduced) when the relative phase is close to constant
(varies rapidly) in time. 
The same tendency is seen in any quartets in resonance 
with different choice of the wave numbers.
Using this result, we can now interpret the time evolution shown in Fig. 3 and 4 in detail. 
Around $t \sim 200$, due to the modulational instability driven by
the finite amplitude parent wave, a pair of side band waves
(daughter waves) appears.
At this stage, phase coherence is generated only within a
small number of resonant quartets, and
the growing sideband waves are restricted within $k<0$ (when $\beta$ is low),
because of the derivative nonlinear term in (11): if a quartet 
includes some of the right-hand polarized wave modes ($k>0$), this 
quartet is "stable" in the sense that the exchange of energy among 
them stays only at a fluctuation level. 
Around $t \sim 300$, the sideband waves are sufficiently large,
and quartets of higher 
harmonic sideband wave modes (including $k>0$ modes) become unstable. 
This broadening of the power spectrum (as seen in Fig. 4) 
corresponds not only to generation of solitary waves (Fig. 3), 
but also to the generation of the phase coherence (Fig. 5), 
since a large number of quartes 
undergo significant nonlinear interaction among them as evidenced
in Fig. 6.  

The interpretation above is justified by inspection of the 
dispersion relation directly computed from the simulation run. 
By computing $\dot{\phi_{k}}$ using (35) and plotting it versus $k_1$, 
the dispersion relation in the simulation system can be visualized for 
any given time. 
In Fig. 7 we show time evolution of the dispersion relation, 
at $t = 0$, $t = 200$, and $t = 550$.
Initially, almost all the wave modes have extremely small amplitude except for
the parent wave, and thus they are located around the linear 
dispersion relation,
$\omega = b_{0}^{2}k+k^2$.
As many of the wave modes acquire finite wave amplitude via the wave-wave
interaction, they tend to be aligned on straight lines (two line 
segments exist in (b): notice a little difference in slopes of lines for 
$k$ positive and negative).
The slope of the line corresponds to the propagation speed of a soliton, and
the interval of $k$ which are aligned on the line segment represent the waves
the soliton is composed of.
At later time (c), it is more evident that the wave modes are concentrated on
a number of different line segments, with different ranges of $k$, and with
different slopes.
These line segments do not necessarily go through the origin, since 
they represent
the {\it four} (but not {\it three}) wave interaction.
Consequently, even though there can be many solitary waves (coherent 
structures)
with different propagation speeds, the resonance condition among the four waves
is well satisfied, \ie, the frequency mismatch is small, and thus a 
large flow of
energy exchange among the modes is expected.

The nonlinear interaction among different waves can be inspected through
evaluation of the so-called bi-coherence (Dudok de Wit \etal, 1999; 
Diamond \etal, 2000),
\begin{equation}
bc(k_1,k_2,k_3) =
\frac{|<\rho_{k1} b_{k2} 
b_{k3}^{*}>|}{\sqrt{<|\rho_{k1}b_{k2}|^2><|b_{k3}|^2>}} .
\end{equation}
Presence of finite wave power at $k_1$, $k_2$, $k_3$, together with $bc\sim 1$
suggests that the waves are in resonance.
Fig. 8 compares time evolution of averaged bi-coherence index, 
$b_{\phi} = N^{-1} \sum bc$, 
total absolute energy flow among all the quartets, 
$F_{tot} = N^{-1} \sum |F(\kk)|$, 
and the phase coherence index, $C_{\phi}$.
On computing $b_{\phi}$ and $F_{tot}$, all the combinations of wave numbers 
($N$ different ways) are exhausted.
Apparently variations of the three quantities agree well to each other, 
suggesting that the phase coherence is generated as the wave energy is
exchanged among resonant quartets. 

%
%
\begin{figure}[bt]
\begin{center}
\includegraphics[width=8cm]{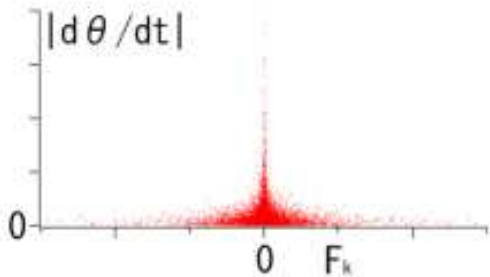}
\end{center}
\caption{Relation between $F(\kk)$ and $|\dot{\theta(\kk)}|$
(see text for detailed explanation of the plot and the variables). 
The set of wave numbers chosen for the plot are 
$k_j =2 \pi m_j / L$, where 
$L=256$ is the system size, and 
$m_3=m_4=m_0=-11$, $m_1=-15$, $m_2=m_3+m_4-m_1 = -7$. 
The figure suggests that 
the exchange of wave energy between the sites is
enhanced (reduced)
when the relative phase is almost constant (varies rapidly) in time.
The unit of the vertical axis is degrees per unit time.
}
\end{figure}
%
%
\begin{figure}[bt]
\begin{center}
\includegraphics[width=8cm]{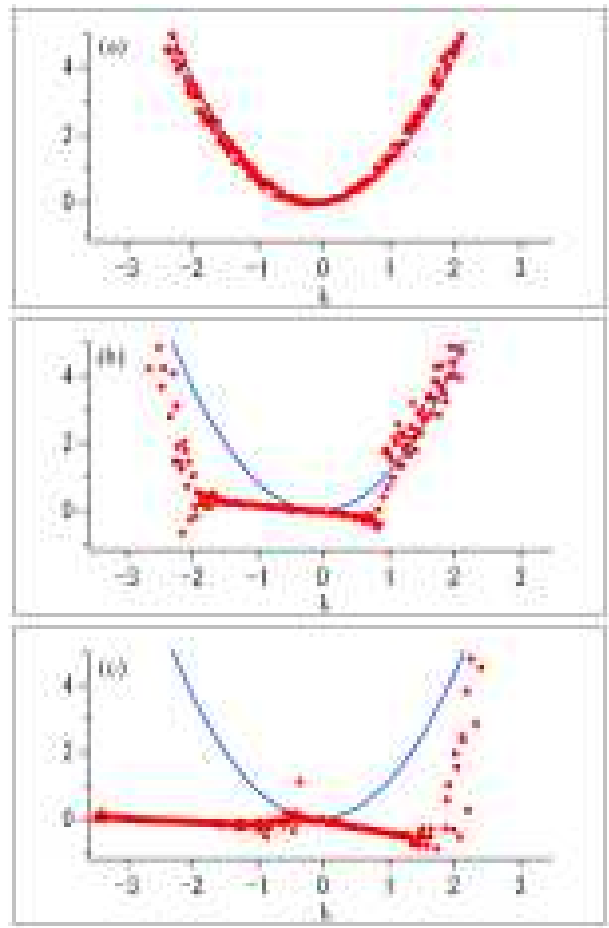}
\end{center}
\caption{
The wave frequency $\dot{\phi_{k}}$ (red circles) versus 
$k$ evaluated at (a) $t=0$, (b) $t=350$, and (c) $t=550$.
At $t=0$, all the wave modes except for the parent wave have very small wave
amplitude, and thus they are located around the linear dispersion relation,
$\omega = b_{0}^{2}k+k^2$ (blue line).
However, as the wave modes grow via wave-wave interaction, they tend to be
aligned along straight lines.
The slope of the line corresponds to the propagation speed of a soliton, and
the interval of k which are aligned on the line segment represent the waves
the soliton is composed of.
Among the quartet of waves on a single line, the frequency mismatch for the
resonance condition $|\dot{\theta(\kk)}|$ is small, and consequently, energy
exchange among them $F(\kk)$ is large, as seen in Fig. 6.}
\end{figure}
%
%
\begin{figure}[bt]
\begin{center}
\includegraphics[width=8cm]{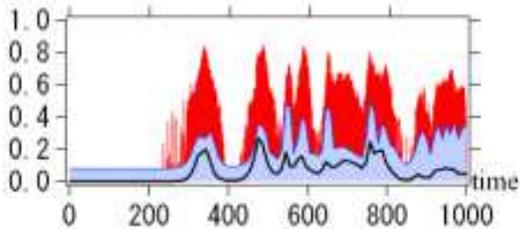}
\end{center}
\caption{
Time evolution of $C_{\phi}$ (red area), $b_{\phi}$ (blue area),
and $F_{tot}/10^{-6}$, (black solid line). 
Variation of $C_{\phi}$ well corresponds to those of $b_{\phi}$ and $F_{tot}$.}
\end{figure}
%
%
\begin{figure}[bt]
\begin{center}
\includegraphics[width=8cm]{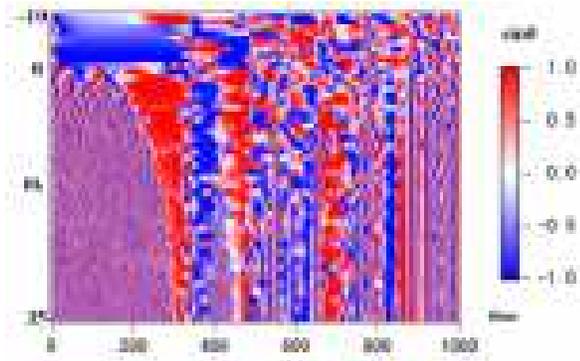}
\end{center}
\caption{Evolution of $\sin\theta(\kk)$,
plotted in the phase space of the wave mode number ($m_1$)
and time.}
\end{figure}

Finally, we make a remark on the direction of the energy flow. 
From Fig. 4 we see that the parent wave mode ($k=k_0=2\pi m_{0}/L$) remains to
be the dominant one throughout the simulation run.
Therefore, among the many quartets of wave modes in resonance
($k_1, k_2, k_3, k_4$) with $k_1 + k_2 = k_3 + k_4$,
the most dominant quartet (which makes (36) the largest) is the one
with $k_3 = k_4 = k_0$.
The resonance condition of wave number for this quartet is,
\begin{equation}
k_0 +k_0 =k_1 + k_2 .
\end{equation}
In fact, the set of wave numbers used for Fig. 6 was chosen so that (38)
is satisfied. 
In Fig. 9 we have plotted $\sin \theta(\kk)$
in the phase space of $m_1$ and time,
with $k_3 = k_4 = k_0$ and $k_2$ determined by (32).
The plot shows the direction of the energy flow between the
parent wave and the daughter waves.
For example, associated with the first formation
of solitary wave rows ($t\sim 200-350$), $\sin \theta(\kk) <0$ for
$k_1 <0$ and $\sin \theta(\kk) >0$ for $k_1 >0$.
Namely, in both regimes of $k_1 > 0$ and $k_1 < 0$,
the energy flow defined in (36) is positive, \ie, the
energy is transferred from the parent wave to the daughter waves.
During the time interval when solitary waves disappear ($t \sim 350-400$),
we observe the opposite, \ie, the energy flows back
from the daughter to the parent waves.

%
%
\section{Nonlinear relation among phase and frequency: Hall-MHD equations}

In this Section, we discuss the modulational instability in the Hall-MHD
equations (Eqs. (5)-(8)), and compare the results with that in the DNLS.

Numerical time integration of the Hall-MHD set of equations, (5)-(8), 
is performed, in a similar way as for the DNLS equation 
discussed in the preceding section.
As initial conditions, finite amplitude, left-hand polarized,
monochromatic, parallel Alfv\'en waves are given, 
with the wave amplitude $b_0=0.4$ and the mode number $m = m_0 = -20$.
Superposed with the parent wave is a small amplitude white noise
with $<|a_{noise}|^2>^{1/2}=10^{-5}$, where $a$ represents any variable in (5)-(8) 
within the range of $-128 < m < 128$. 
The system size is $L=256$, in which there are 1024 grid points.
We chose $\beta=0.5$  
so that the modulational instability growth rate is the
largest among all the parametric instabilities possibly driven unstable.
The parameters specified here are essentially equivalent to the ones 
used in the previous section. 

Fig. 10 shows time evolution of $|b|$ plotted in a way  similar to Fig. 3, 
and Fig. 11 represent time evolution of $\log|b_m|$ and $\log|\rho_m|$,
which is the power of the transverse and the longitudinal waves, respectively,
as a function of the wave mode number, $m$.

%
%
\begin{figure}[bt]
\begin{center}
\includegraphics[width=8cm]{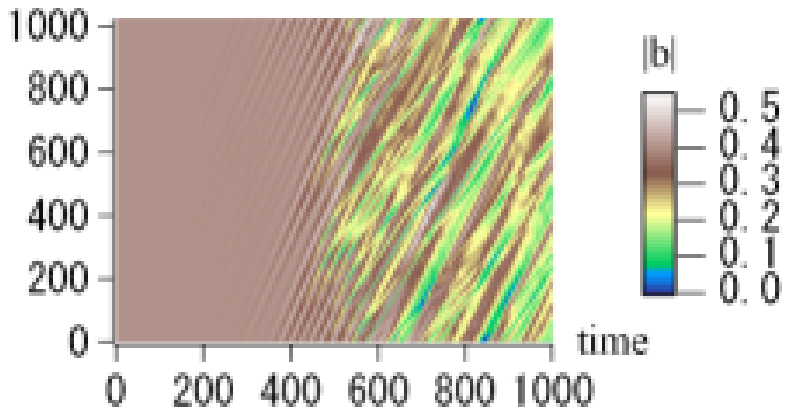}
\end{center}
\caption{Time evolution of envelope $|b|$ due to Eqs.(5-8) with periodic
boudary conditions. Initial conditions are given as a superposition
of finite amplitude, left-hand polarized monochromatic waves,
and very small amplitude white noise.}
\end{figure}
%
%
\begin{figure}[bt]
\begin{center}
\includegraphics[width=8cm]{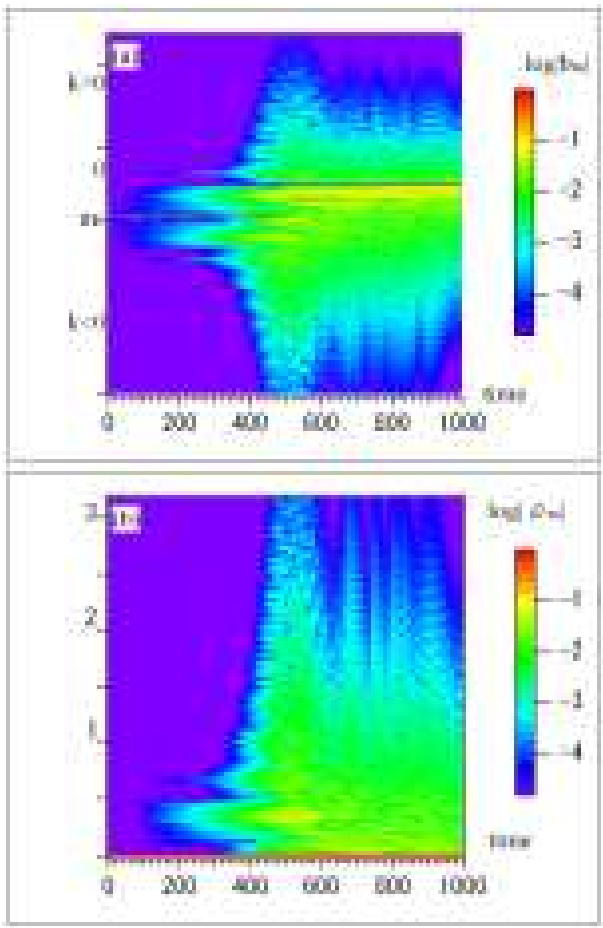}
\end{center}
\caption{Time evolution of power spectrum (in logarithmic scale),
$\log|b_m|$ and $\log|\rho_{m}|$, plotted in the phase space of the wave mode
number ($m$) and time.
The parent mode is given initially at $m_0=-20$.
Positive (negative) $m$ corresponds to positive (negative) helicity.
}
\end{figure}
%
%
\begin{figure}[bt]
\begin{center}
\includegraphics[width=6cm]{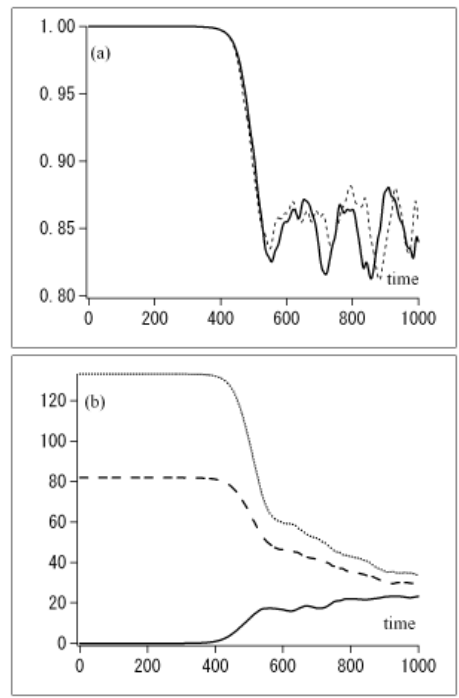}
\end{center}
\caption{Time evolution of (a) the cross-correlation, $c (\lambda=0, \tau=0)$ 
evaluated for the run shown in Fig. 10 (solid line),
and $c(0,0)$ evaluated for a run in which the advection term is 
artificially removed from the Hall-MHD set of equations (broken line). 
(b) magnetic field energy (broken), perpendicular kinetic energy (dotted), 
and parallel kinetic energy (solid), plotted versus time. 
Around $t \sim 350$, together with the decay of the 
cross-correlation function, 
the magnetic field and the perpendicular kinetic energy
decrease and the parallel kinetic energy increases. 
}
\end{figure}

From comparison of the DNLS and the Hall-MHD simulation runs, we dicuss validity
of various approximations assumed in the DNLS. 
In particular, we look at the following:
(a) the quasi-static approximation, (10),
(b) conservation of the magnetic field energy alone, and
(c) the assumption of constant plasma density in the dispersion term
(\ie, the Hall-term, ${\partial/ \partial x}(\rho^{-1} \partial b/ \partial x)$,
is simplified by assuming $\rho$ to be constant). 
These approximations are justified as long as the power of the
longitudinal modes (Fig. 11(b)) is much less than that of the
transverse modes (Fig. 11(a)).
Around $t \sim 350$, the power spectrum of sideband wave modes and
the daughter wave modes begin to grow.
To check the validity of the approximation in (10), we evaluate the 
cross-correlation function,
\begin{equation}
c(\lambda,\tau)=\frac{<|b(x+\lambda,t+\tau)|^2 \rho(x,t)>}
{\sqrt{<|b(x+\lambda,t+\tau)|^4><(\rho(x,t))^2>}} 
\end{equation}
where we simply let $\lambda=\tau=0$ in the present analysis.
The result is plotted as a solid line in Fig. 12(a).
When $0<t< \sim 350$, relation (10) is well held, but
around $t \sim 350$, the cross correlation starts to decrease.
Fig. 12(b) shows time evolution of the magnetic field energy (broken line),
parallel kinetic energy (solid line),
and perpendicular kinetic energy (dotted line).
The decay of the correlation function occurs at the same time
as the magnetic field and the perpendicular kinetic energy both decrease.
Around $t \sim 500$, inverse cascade begins to take place 
(discussion of the inverse cascade in the DNLS can be found 
in Krishan and Nocera, 2003).

The validity of the static approximation has been examined by several authors.
By performing hybrid (kinetic ions + an electron fluid) simulations, 
Machida \etal (1987) argues that the system automatically adjusts itself to
a state described by the static approximation when ion kinetic effects are included. 
However, as time elapses, longitudinal mode grows and
the static approximation is soon violated.
Spangler (1987) concluded that 
the breakdown of the static approximation takes place in association with rapid evolution
of wave packets by ponderomotive force.
Combining (5) and (6), we have
\begin{equation}
\frac{\partial^{2} \rho}{\partial t^{2}}-\beta\frac{\partial^{2}
\rho}{\partial x^{2}} = \frac{\partial^{2}}{\partial x^{2}}\bigr(\frac{
|b| ^{2}}{2}+\rho u^{2}\bigr) . 
\end{equation}
This equation describes
propagation of sound waves in a presence of
a source term (r.h.s.), which consists of 
the ponderomotive force, $|b|^2/2$, 
and the advection effect, $\rho u^2$ (Spangler, 1987).
The latter represents the "self-coupling" of longitudinal waves.

In order to clarify the roles of the "self-coupling" term,
which is not discussed in the analysis of Spangler (1987),
we have run numerical experiment in which the advection terms are
artificially removed from the Hall-MHD set of equations.
The broken line in Fig. 12(a) depicts time evolution of the
cross correlation function evaluated for the run without the advection terms,
using the same simulation parameters as the previous run.
The time evolution for both runs are essentially the same, suggesting
that the process playing the most dominant role in violating the
static approximation is the ponderomotive force, as suggested by
Spangler (1987).
The self-coupling term cannot be neglected, however, since its magnitude
becomes comparable to other terms as the longitudinal wave modes grow.

In a similar way described before to plot 
the dispersion relation of nonlinear Alfv\'en waves
in the DNLS directly from the simulation run, we have computed the 
$\dot{\phi}_k^{\rho}$ and $\dot{\phi}_k^{b}$, using (5) and (8), respectively.
Corresponding to appearance of solitary waves and localized structures,
the frequency distribution ($\dot{\phi}_k^{\rho}$ and $\dot{\phi}_k^{b}$),
\ie, the dispersion relation,
appears along a straight line, as shown in numerically evaluated
dispersion relation, Fig. 13.
Since these waves propagate in the same direction as the parent wave,
the frequency distribution along a straight line is usually satisfied
at the origin.
Furthermore, Fig. 13 shows that the longitudinal
and transverse wave modes have a similar slope, which is the velocity of
the structures composed of these waves.
Thus, the three wave resonance condition between the transverse
wave modes and the longitudinal wave modes is approximately satisified.

Fig. 14 displays time evolution of the wave frequency
($\dot{\phi}_k^{b}$) for the parent wave ($m=-20$, solid line),
and some of the daughter waves ($m=-8$, dotted line)
($m=-3$, broken line).
When $0 < t <\sim 500 $, the parent wave mode is dominant,
and its frequency stays almost constant.
Around $t \sim 500$, the inverse cascade begins to take place, and
the parent wave frequency starts to be modified.
On the other hand, the frequencies of the lower sideband wave modes
remain almost constant (broken line).
Thus, the phase velocity of the modes with
the maximum power is close to the velocity of the structures,
and it corresponds to the slope of the dispersion relation in Fig. 13.

Now we examine the approximation (c) we listed before in the DNLS, 
\ie, the assumption of constant density in the Hall term. 
In order to do so, let us explicitly write down the phase relation
of (8), 
\begin{equation}
\frac{d \phi_{k}^{b}}{d t} =
\omega_{nl}+\omega_{disp}+\omega_{lin}  ,
\end{equation}
\begin{equation}
\omega_{nl}=
k\sum_{k=k1+k2}\frac{|u_{k1}||b_{k2}|}{|b_{k}|}
\cos(\theta_{k=k1+k2}^{bub}) ,
\end{equation}
\begin{equation}
\omega_{disp}=k_{3}k\sum_{k=k3+k4}
\frac{|b_{k3}||V_{k4}|}{|b_{k}|}
\cos(\theta_{k=k3+k4}^{bbV}) ,
\end{equation}
\begin{equation}
\omega_{lin}=-k\frac{|v_{k}|}{|b_{k}|}\cos(\phi_{k}^{b}-\phi_{k}^{v}) ,
\end{equation}
where $V=\rho^{-1}$ and $\theta_{k=l+m}^{xyz}=\phi_{k}^{x}-\phi_{l}^{y}-\phi_{m}^{z}$.
In the linear perturbation analysis in section 2.2 we have called
$\omega_{nl}$, $\omega_{disp}$, and $\omega_{lin}$ as contribution to the
wave frequency via 'nonlinear', 'dispersion', and 'linear' effects, 
respectively.
The dispersion effect is not simply $k^2$, but is 
$k^2 - b_0 k_0 K \rho/2/b_{\pm}^{(*)}$, 
where the latter term arises due to longitudinal perturbation. 
Just like in the DNLS model, the balance between the
'nonlinear' and the 'dispersion' terms also exist
in the Hall-MHD equations.

Fig. 15 shows $\omega_{nl}$, $\omega_{disp}$, and $\omega_{lin}$
evaluated at $t$=350, 500, and 720. 
We see that $\omega_{lin}$ is linear in $k$: this may be regarded
as self-generation of the (dispersive) Walen's relation, 
which is the relation between the transverse magnetic and the 
velocity field satisfied for Alfv\'enic fluctuations and 
rotational discontinuities ({\it e.g.}, Landau and Lifshitz, 1962). 
For example, the normalized magnetic and the velocity field, 
$b_0$ and $v_0$ in (14-15), satisfy 
$v_0=\mp b_0/v_{\phi 0}$, where the signs represent parallel/anti-parallel
propagation with respect to the background magnetic field, 
and the phase velocity $v_{\phi 0}$ is determined from the 
dispersion relation (16). 
The term $\cos (\phi_{k}^{b}-\phi_{k}^{v})$ in the r.h.s. of (44) 
takes the value of either 1 or -1 (so that $\omega_{lin}/k =\pm|v_{k}|/|b_{k}|$)
in association with broadening of the power spectrum. 
Also, from Fig. 13 and 15, we find numerically that 
$\omega_{lin}/k \sim k/\omega$ (corresponding to the appearance of
near straight lines in these plots). 
Therefore we find that the system approaches automatically to a state
where $k/\omega =\pm|v_{k}|/|b_{k}|$
(Walen's relation) is automatically fulfilled.
Furthermore, since $\omega_{lin}$ is almost linear in $k$
in Fig. 13, so is the sum of $\omega_{disp}$ and $\omega_{nl}$
as (41) suggests, 
\ie, the contribution to the wave frequency nonlinear in $k$ is 
cancelled due to balance between the dispersion and the nonlinear effects.
%
%
\begin{figure}[hbt]
\begin{center}
\includegraphics[width=8cm]{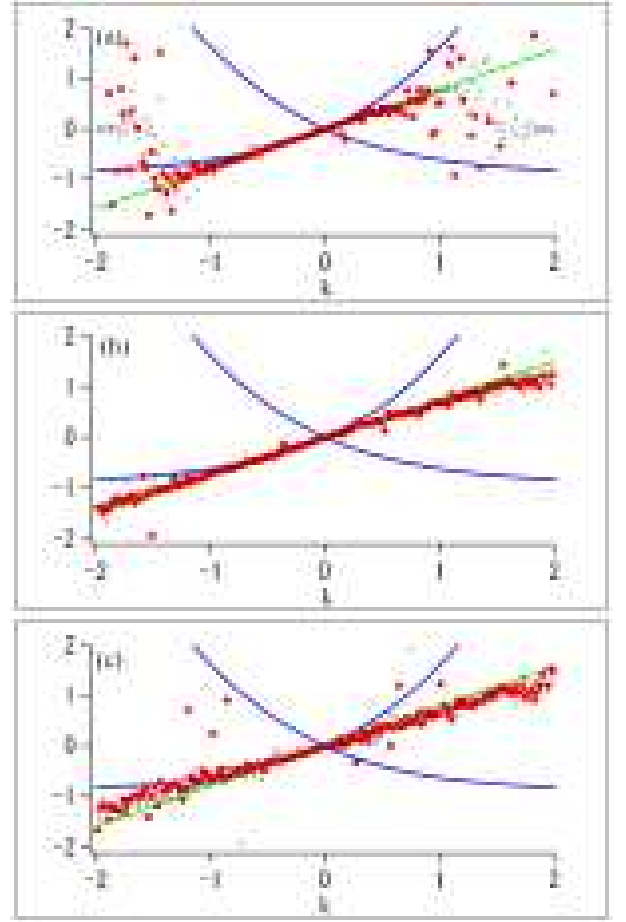}
\end{center}
\caption{Numerically obtained dispersion relation at 
(a) t=350, (b) t=500, (c) t=720.
The blue line indicates the
linear dispersion relation, and the red circles and black crosses show the
frequency evaluated from the temporal variation of the magnetic field 
mode ($\dot{\phi}_k^{b}$)
and that of the density modes ($\dot{\phi}_k^{\rho}$), respectively.
Superposed (green) line has a slope equal to the phase velocity
of the wave mode which has the maximum power.
Corresponding to the appearance of solitary waves and
localized structures, the wave power tends to be aligned along a
straight line. 
}
\end{figure}
%
%
\begin{figure}[hbt]
\begin{center}
\includegraphics[width=8cm]{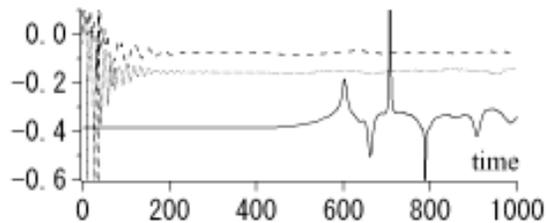}
\end{center}
\caption{
Time evolution of the wave frequency ($\dot{\phi}_k^{b}$) 
for the waves
with $m=-20$ (solid line), $m=-8$ (dotted line), and
$m=-3$ (broken line).
The wave number of these modes are $k=-0.49$, -0.19, and -0.07, respectively.
When $0 < t \lesssim 500 $, the parent wave mode is dominant,
and its frequency is almost constant.
Around $t \sim 500$, the inverse cascade begins to take place,
and the parent wave frequency starts to vary.
On the other hand, the frequencies of the lower sideband wave modes
remain almost constant after $t>500$.}
\end{figure}
%
%
\begin{figure}[hbt]
\begin{center}
\includegraphics[width=8cm]{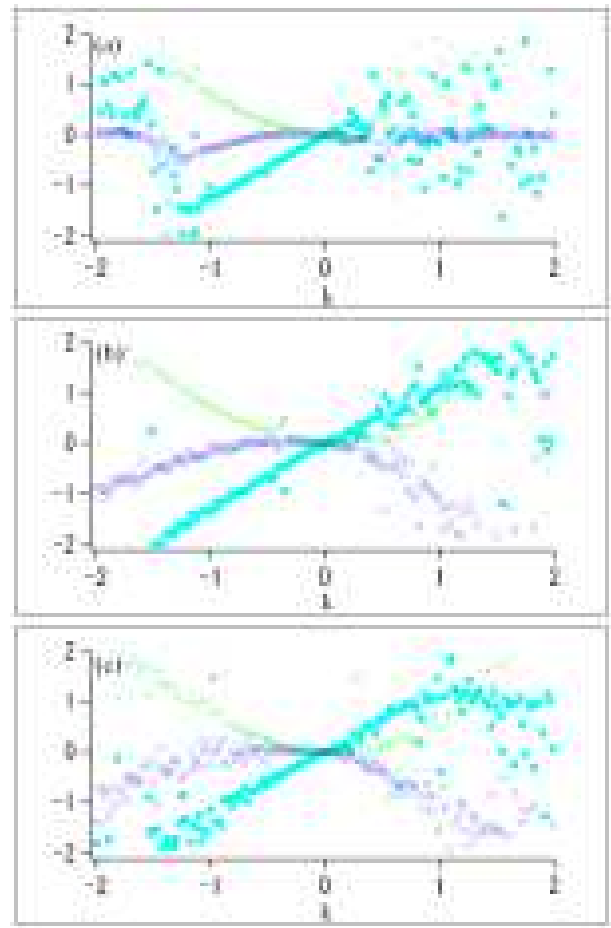}
\end{center}
\caption{
The three basic elements of the wave frequency, 
$\omega_{lin}$ (light blue squares), 
$\omega_{nl}$ (blue circles), and 
$\omega_{disp}$ (green crosses), 
defined in (42)-(44), plotted versus $k$, 
evaluated at (a) $t=350$, (b) $t=500$, and (c) $t=720$. 
As discussed in Section 2.2, $\omega_{lin}$ is almost linear in $k$, 
suggesting that the dispersive Walen's relation is held. 
Since the dispersion relation (Fig. 13) suggests that the 
wave frequency, which is a sum of the three terms above, is linear
in $k$ also, the linearity in $k$ of the sum of 
$\omega_{disp}$ and $\omega_{nl}$ should be held as well, 
indicating that the dispersion and the nonlinear effects cancel 
each other, and the system is brought to a state where the wave
resonance condition is easily satisfied.
}
\end{figure}

Fig. 16 shows distribution of $\dot{\phi}_k^{u}$ (similar
to $\dot{\phi}_k^{\rho}$) versus $k$.
When $0 < t \lesssim 500 $, these longitudinal fluctuations are 
found to be distributed along the same straight line in the
dispersion relation, and should be recognized as 
the ponderomotive density fluctuations, 
produced via interaction of Alfv\'en waves.
In the present case, the fluctuations are produced
in the course of the modulational instability, 
and the two interacting Alfv\'en waves are 
on the same branch with nearly equal wave numbers and the wave 
frequencies, so that the resultant ponderomotive fluctuations 
have similar phase velocity as the interacting Alfv\'en waves.
The phase velocities of the longitudinal fluctuations 
($\rho$ and $u$ in Fig. 16), 
transverse fluctuations ($b$ and $v$ in Fig. 13), 
and the acoustic wave (solid line in Fig. 13), are 
measured as 
$\sim 0.78$, $\sim 0.78$, and $\sim 0.70$, respectively. 
The former two are (almost) equal, and are distinct from the last one. 
Thus we conclude the longitudinal fluctuations observed at 
$0 < t \lesssim 500 $ are 
the ponderomotive density fluctuations, rather than the sound wave.
Around 
$t \sim 500$, however, the dispersion relations of the longitudinal 
fluctuaions start to approach the sound wave branch.
This is a consequence of the nonlinear development of the modulational
instability, in which the ponderomotive density fluctuation acts as
a driver to excite the sound waves, as represented in (40).

This excitation of the sound wave is also related to particle heating
due to the modulational instability of Alfv\'en waves.
Many authors discussed Landau damping associated with envelope
modulation of Alfv\'en waves
(Mj{\o}lhus and Wyller, 1986,1988; Spangler; 1989, 1990;
Medvedev and Diamond, 1996; Passot and Sulem; 2003).
However, 
the ion acoustic wave mode is exclueded, as long as the "ponderomotive" expansion
is used in scaling of the system. 
In the case of the weak nonlinearity, the "triple-degenerate" expansion makes
a similar description at $\beta \sim 1$ (Hada, 1993).
The kinetic "triple-degenerate" DNLS equation will be useful for
elucidating the physics of the heating process.
The use of numerical simulation models (Vasquez, 1995) and/or
theoretical models (Snyder, 1997; Passot and Sulem, 2004) that include the 
strong nonlinearity and kinetic effects are desired.

%
%
\begin{figure}[hbt]
\begin{center}
\includegraphics[width=8cm]{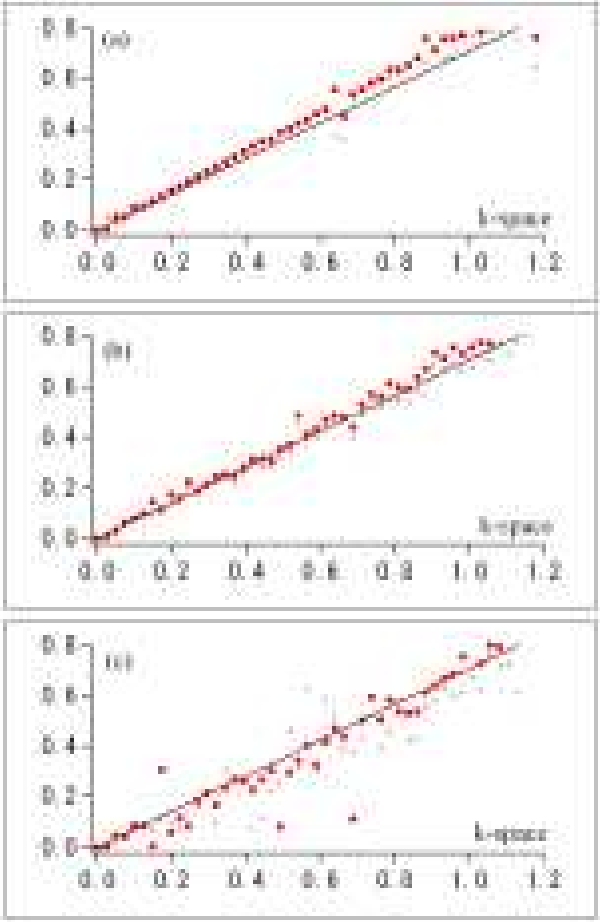}
\end{center}
\caption{
Numerically computed frequency $\dot{\phi}_k^{u}$(red circles) and 
$\dot{\phi}_k^{\rho}$ (black crosses) versus the wave number $k$ at
(a) t=350, (b) t=500, (c) t=720. 
When $0 < t \lesssim 500 $, the longitudinal
modes $\dot{\phi}_k^{u}$, $\dot{\phi}_k^{\rho}$ are distributed 
around a line slightly (but distinctly) above the sound wave
branch (solid black line). 
The former may be called the "ponderomotive density fluctuation"
branch. 
As later times, however, we find that 
the longitudinal wave frequencies are re-distributed
around the sound branch. }
\end{figure}

%
%
\section{Conclusions and discussions}

In this paper, we discussed the nonlinear relation among phases and
frequencies in the modulational instability of parallel propagating 
Alfv\'en waves
in the context of one-dimensional two fluid models, using linear perturbation
analysis and numerical experiments.

The main results of the present paper may be summarized as follows:

(1) Corresponding to generation of solitary waves or
coherent structures by the wave modulation, dispersion relation
of the waves tend to be aligned along straight
segments of lines (Fig. 7 and Fig. 13).
The slopes of these straight lines correspond to velocities of
the solitary waves and the structures, and are well estimated as the
phase ($\sim$ the group) velocity of the wave with the largest wave power.

(2) The exchange of wave energy
among the resonant wave modes is enhanced (reduced) when
the relative phase is close to constant (varies rapidly) in time (Fig. 6).
This is a universal feature in a wide variety of nonlinear systems: 
{\it e.g.,} it can also be seen in a single triplet model (Appendix).

(3) The fact that the dispersion relation is essentially
given as segments of lines in the dispersion relation suggests that
the mismatch in the resonance conditions among the coupled wave modes
is reduced automatically.
The waves can efficiently exchange wave energy and generate the
phase coherence.

In the DNLS equation (Section 3), increase and decrease of the phase
coherence index well corresponds to the appearance and disappearance
of solitary wave trains.
However, in the Hall-MHD equations (Section 4),
evolution of the phase coherence index turns out to be more complex.
There are two issues to be discussed: the physical processes leading to
the variation of the phase coherence index, and the validity of the index.

In the DNLS, the energy flow among the wave modes
generates order not only among the wave frequencies
but also among wave phases (see Appendix, and also Fig. 9), 
and the associated broadening of the power spectrum
always corresponds to the global energy flow because of the balance
between the dispersion effect and the nonlinear effect.
However, in the Hall-MHD, 
the presence of the localized structures does not always correspond 
to the broadening of the power spectrum.
In Fig. 17 we plot time evolution of $C_{\phi}$ and $b_{\phi}$ based on 
the Hall-MHD simulation run discussed before. 
The broadening of the power spectrum correlates well with $b_{\phi}$, 
but not quite well with $C_{\phi}$.
This probably is due to the fact that the 
structures observed in the simulation are not always generated by 
the local energy flow. 

In case of the modulational instability starting from a single
parent wave plus small amplitude background noise, 
one can evaluate the relative phase among the resonant wave modes
and also the energy flow among them, 
and identify clearly time intervals the phase coherence is generated
(see Fig. 9, and also early part of Fig. 18). 
On the other hand, after the initial stage of the modulational 
instability, wave modes at low wave numbers produced via the
inverse cascade become dominant, it is no longer possible to 
identify the time intervals in which the phase coherence is produced. 

Regarding the phase coherence index, we have to say that 
its physical meaning is still not completely clear, 
although it was demonstrated that the index 
is a useful measure to identify the 
presence of wave-wave interaction (Koga and Hada, 2003). 
Since the concept of the phase synchronization is universal
({\it e.g.}, He and Chian, 2005), it is important to develop a method that
can characterize the underlying nonlinearity of the system.
This discussion is translated into the explanation or
improvement of the phase coherence index.

The relation between the
longitudinal and transverse modes has important information on
the nature of the parametric instability. 
In the dispersion relation, 
longitudinal and transverse modes were found to be both linear,
with almost the same slope, corresponding
to the structures generated by modulational instability. 
On the other hand, the longitudinal fluctuations generated by the 
decay instability distribute on the sound wave line with 
some band width (Terasawa \etal, 1986). 
Such a cross relation among different variables in MHD turbulence is 
very important to understand the turbulent dynamics.
Henceforth, we have to analyze in detail the general rules of the
relation among these variables in parallel Alfv\'en turbulence generated by
parametric instability. 
Also, it is essential to include the kinetic effects when $\beta$ is
moderately large. 
%
%
\begin{figure}[bt]
\begin{center}
\includegraphics[width=8cm]{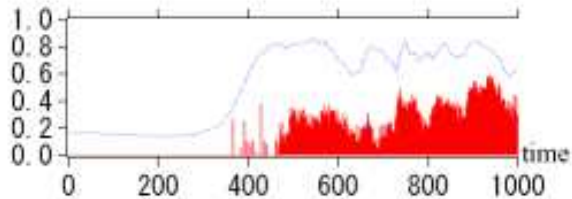}
\end{center}
\caption{Time evolution of $C_{\phi}$ (red area) and $b_{\phi}$ (broken line). 
In association with the broadening of the power spectrum, 
$b_{\phi}$ varies accordingly, but $C_{\phi}$ does not. 
}
\end{figure}
%
%
\begin{figure}[hbt]
\begin{center}
\includegraphics[width=8cm]{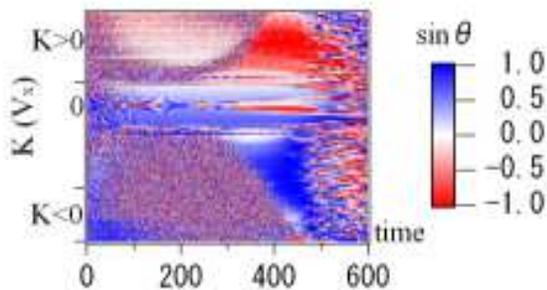}
\end{center}
\caption{
Time evolution of the relative phase between the parent
and the other wave modes, 
$\sin (\phi_{k0}^{b} - \phi_{k1}^{b}-\phi_{K}^{u})$.
}
\end{figure}

It is important to identify the phase coherence brought by nonlinear wave-wave
interaction, in order to discuss underlying physical processes leading to the
generation of the coherent structures.
This is particularly important for analysis of spacecraft data, since
there are many localized structures in space plasmas
which are produced, presumably, by processes other than nonlinear 
wave-wave interactions. 
Dudok de Wit \etal (1999) and Soucek \etal (2003) discussed the so-called
Volterra model, a way to identify nonlinear wave-wave interactions 
using the satellite and the simulation data, by utilizing
the high order spectrum analyses and weak turbulence theory.
This method may shed light on understanding the nature of wave-wave 
interactions when the weak and strong turbulence processes
co-exist. 
In the future, we hope to clarify the process of the phase coherence generation
via the wave-wave interaction in more detail, and to extend the 
Volterra-type models in such a way that the weak turbulence approximation
is not necessary. 

Finally, we make a remark on energetic particle transport
in the turbulence including the localized structures.
Since these structures have broad power spectrum in the
$k$ space, particles with wide range of energy can resonate with them.
However, this is not the more important side of the
wave-particle interaction, from the viewpoint of the
phase coherence -- rather, it is in the fact
that large amount of electromagnetic field and plasma energy is
concentrated within the solitary wave (phase correlated wave/structure),
so that it can influence particle motion via strong and correlated
impulse force rather than via sequence of random forces which
last for a long time.

For example, let us consider pitch angle diffusion due to
the turbulence consisting of superposition of parallel,
circularly polarized Alfv\'en waves (slab model). Within this model,
it is well known that the particles cannot diffuse across 90 degrees
pitch angle, within the framework of the quasi-linear diffusion,
simply because there are no waves which can resonate with
particles without parallel velocities. However, the mirror force can reflect
the particles quite easily, if there exist solitary waves with a finite
amplitude variation of the total magnetic field. Furthermore, in a presence
of many of these solitary waves, individual particle trajectory may
be given
as combination of quasi-ballistic motion between the solitary waves and
trapping by one of the solitary waves. Fermi type acceleration is possible
(Kuramitsu and Hada, 2000), and furthermore, depending on relative dominance
between the ballistic and trapped trajectories, ensemble of these particles
may appear
as either super- or sub-diffusive (Zimbardo \etal, 2000; Carreras \etal,
2001).
In order to properly describe time evolution of the ensemble, one has to
invoke the fractal diffusion formalism (Metzler and Nonnenmacher, 1998;
del-Castillo-Negrete \etal, 2004).
%
%
\begin{acknowledgements}
We thank Drs. S. Matsukiyo and D. Koga for fluitful discussions and valuable comments. 
This paper has been supported by JSPS Research Felowships for Young Scientists in Japan.
\end{acknowledgements}

%
%
\section*{Appendix: Three wave resonance: A single triplet}

A triplet is known as the most basic element of nonlinear interaction that
can be extracted from various phenomena.
Decay instability of Alfv\'en waves in space plasma is an example.
A finite amplitude Alfv\'en wave propagating parallel to the ambient
magnetic field decays into a backward propagating Alfv\'en wave and a
forward propagating ion acoustic wave.
We consider a set of three wave equations (3W) discussed by 
Sagdeev and Galeev (1969),
\begin{eqnarray}
\dot{C_1} &=& -iC^*_2C_3  ,\\
\dot{C_2} &=& -iC^*_1C_3  ,\\
\dot{C_3} &=& -iC_1C_2  ,
\end{eqnarray}
where $C_j$ is the normalized complex amplitude of the $j$th mode.
The above set of equations is derived via resonance condition 
among frequencies of each mode, 
\begin{eqnarray}
\omega_3=\omega_1+\omega_2  .
\end{eqnarray}
By introducing the wave quanta (wave action), 
$N_j = |C_j|^2 = \epsilon_j /\omega_j$, where $\epsilon_j$ is the 
wave energy of mode $j$, we have the Manley-Rowe relations, 
\begin{eqnarray}
N_1+N_3 &=& const ,
\end{eqnarray}
\begin{eqnarray}
N_1-N_2 &=& const  .
\end{eqnarray}
Now let us write $C_j=A_j\exp{(i\phi_j)}$, 
with $A_j(=\sqrt{N_j})$ and $\phi_j$ real, 
then we obtain a relation about the phase difference among the triplets 
($\theta=\phi_3-\phi_2-\phi_1$),
\begin{eqnarray}
\dot{\theta} &=& \dot{\phi_3}-\dot{\phi_2}-\dot{\phi_1}  ,\\
&=& -(\frac{A_1A_2}{A_3}-\frac{A_2A_3}{A_1}-\frac{A_3A_1}{A_2})\cos{\theta}  ,\\
&=& \cot{\theta}\frac{d}{dt}\log(A_1A_2A_3)  ,
\end{eqnarray}
where $\dot{\phi_j}$ is the (nonlinear) frequency and
$\dot{\theta}$ represents the frequency mismatch in the resonance condition.
From the above, the following conservation law is derived
\begin{eqnarray}
A_1A_2A_3\cos\theta = D_0 =const .
\end{eqnarray}

The 3W equations at a glance have six degrees of freedom,
but if we write them into the real amplitude and phase, we find that 
the equations
actually have only four degrees of freedom: the real amplitudes
$A_1$, $A_2$, $A_3$,
and the phase difference $\theta$.
Since there are three invariants ((49), (50), (54)),
the 3W equation set is integrable. 
All initial problem of 3W can be solved, and the solution 
can be written using elliptic functions.
Behavior of the system may be physically interpreted as either stable 
or unstable depending on initial distribution of wave quanta to 
the three eigenmodes (Fig. 19 and 20).
%
%
\begin{figure}[bt]
\begin{center}
\includegraphics[width=8cm]{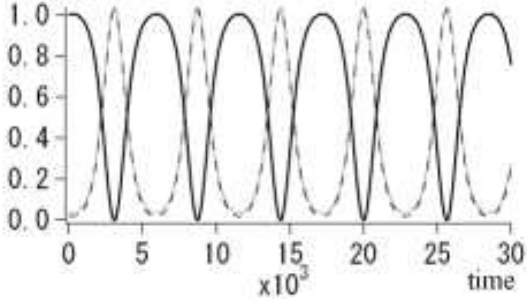}
\end{center}
\caption{Time evolution of $N_j$ with the initial condition,
$N_3 >> N_1, N_2$. 
The broken, dotted, and solid lines indicate $N_1$, 
$N_2$, and $N_3$, respectively.
The system is 'unstable' in that the 'quanta' originally stored in 
mode 3 are re-distributed
to modes 1 and 2 within an instability time scale
($\sim 3000$). }
\end{figure}
%
%
\begin{figure}[bt]
\begin{center}
\includegraphics[width=8cm]{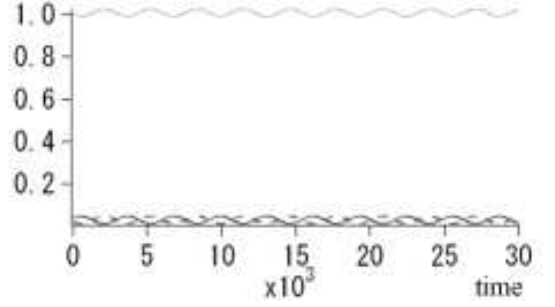}
\end{center}
\caption{Same as Fig. 19 except that the initial condition
here is $N_2 >> N_1, N_3$.
The broken, dotted, and solid lines indicate $N_1$, $N_2$, and $N_3$, respectively. 
The system is 'stable'
in the sense that variations of $N_1$, $N_2$, and $N_3$
are small.}
\end{figure}

Let us define the flow of quanta $F$, which represents the interaction 
between the eigenmodes,
\begin{eqnarray}
F=\dot{N_1}=\dot{N_2}=-\dot{N_3}=2A_{3}A_{2}A_1\sin{\theta} ,
\end{eqnarray}
and consider its relationship 
to temporal change of the phase difference $\dot{\theta}$.
The sign of $\sin\theta$ determines the direction of
quanta flow. 

As for the initial conditions there are two distinct possibilities.
The first is that there are some modes without any quanta, 
and the second is that all the modes have some finite number of quanta.
In the first case, $\cos\theta = 0$ always since $D_0 = 0$.
In Fig. 21, we see that $\theta$ jumps $180$ degrees periodically
because the direction of the quanta flow remain unchanged per half the 
oscillation period.

%
%
\begin{figure}[bt]
\begin{center}
\includegraphics[width=8cm]{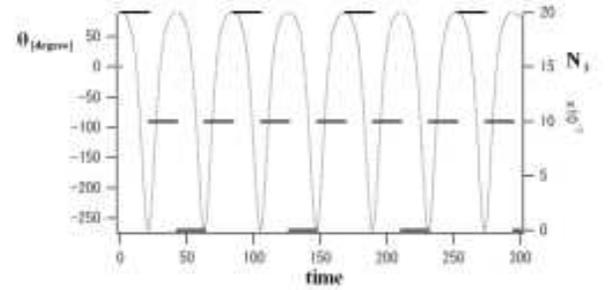}
\end{center}
\caption{Time evolution of $N_1$, $N_2$, $N_3$, and $\theta$
for the case that $A_1A_2A_3\cos\theta=0$.}
\end{figure}

From now on, we only discuss the case where all the modes have some 
finite number of quanta.
Since $D_0=const\neq0$,
we can rewrite $F$ as
\begin{eqnarray}
F=2D_0\tan{\theta} ,
\end{eqnarray}
and therefore, the condition $-\pi/2<\theta<\pi/2$ is required in order
that $F$ be finite.
This condition is evident from (54).
Eq. (56) suggests that $|F|$ is large 
when $\theta$ is close to either $-\pi/2$ or $\pi/2$,
while $F=0$ when $\theta=0$ (Fig. 22(a)).
%
%
\begin{figure}[bt]
\begin{center}
\includegraphics[width=8cm]{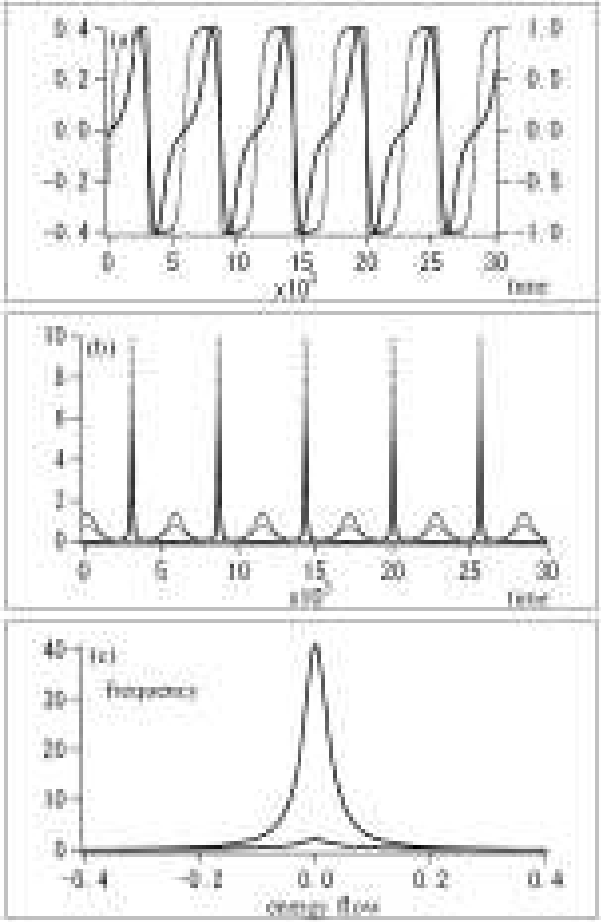}
\end{center}
\caption{(a) The energy flow $F$ (solid line, left axis) and
$\sin \theta$ (dotted line, right axis) are plotted versus time.
When the energy is exchanged between the modes, the relative 
phase stays at almost at a constant level. 
The same tendency is observed in systems with multiple 
degrees of freedom ({\it cf.}, Fig. 9 and 18). 
(b) Time evolution of 
$\dot{\phi_j}$. 
The thin, solid, and dotted lines indicate 
$\dot{\phi_1}$, $\dot{\phi_2}$, and $\dot{\phi_3}$, respectively. 
(c) Relation between $\dot{\theta}$ and $F$ for a (single) triplet. The 
exchange of quanta between the sites is enhanced when the relative 
phase is approximately constant in time, while the exchange of quanta 
is reduced when the relative phase varies rapidly. }
\end{figure}

Fig. 22(b) shows time evolution of $\dot{\phi_j}$.
Both fast and slow temporal changes of $\theta$ can be seen.
Around the point where the number quanta is close to take extremal values,
the temporal change of $\theta$ becomes rapid.
If the system is unstable, temporal change of the phase difference is 
much faster when the quanta is at local minimum, than when it is at
local maximum. 
Fig. 22(c) shows this relationship more clearly.
Using (54) and (56), and $\dot{\phi_j}=-D_{0}/N_{j}$, we have
\begin{eqnarray}
\dot{\theta}=-D_0(\frac{1}{N_3}-\frac{1}{N_1}-\frac{1}{N_2})=D_0\frac{\dot{F}}{2G^2} ,
\end{eqnarray}
where $G = A_1A_2A_3$.
This equation exactly shows the relationship discussed above.

In summary, if the temporal change of the phase difference is small (large),
quanta flow between the modes is large (small).
In other words, the stronger (weaker) the modes are correlated,
are the interaction between the modes strong (weak).
This relationship between the phase coherence and the
interaction among eigenmodes is universal.
For example, similar relationship is held in a system where many 
triplets are connected, in the DNLS, and in the Hall-MHD models.
Also, when $\cos\theta =0$ initially, 
we have a situation similar to the case that $N_j = 0$ for 
one (or more) of the modes, and thus $\cos\theta =0$ always.


\end{document}